\documentclass[final,authoryear,3p,times]{elsarticle}

\usepackage{amsthm,amsmath} 
\usepackage{amssymb} 
\usepackage{algorithm} 
\usepackage{algorithmic}
\usepackage{pstool}
\usepackage[caption=false]{subfig}
\usepackage[colorlinks,citecolor=blue,urlcolor=blue]{hyperref} 
\usepackage{hypernat}

\def\F{{\mathcal F}}

\def\argmax{\mathop{\rm argmax}} 
\def\argmin{\mathop{\rm argmin}}

\newcommand{\s}{\ensuremath{\mathbb{S}}} 
\newcommand{\real}{\mathbb{R}} 
\newcommand{\fspace}{\ensuremath{\mathcal{F}}} 
\newcommand{\T}{\ensuremath{\mathsf{T}}} 
 
\newcommand{\ltwo}{\ensuremath{\mathbb{L}^2}} 
 
\newcommand{\inner}[2]{\left\langle #1,#2 \right\rangle}

 \numberwithin{equation}{section}

\newtheorem{definition}{Definition} 
 \theoremstyle{plain} 

\hypersetup{pdfauthor = {J. Derek Tucker, Wei Wu, and Anuj Srivastava}, pdftitle = {Generative Models for Functional Data using Phase and Amplitude Separation}, pdfsubject = {Computational Statistics}, pdfkeywords = {amplitude variability, function alignment, unction principal component analysis, functional data analysis, generative model, phase variability}, pdfcreator = {LaTeX with hyperref package}, pdfproducer = {pdflatex}}

\journal{Computational Statistics \& Data Analysis}
\begin{document} 

\tabcolsep 2pt 
\begin{frontmatter}
\title{Generative Models for Functional Data using Phase and Amplitude Separation}

\author[nswc,fsu]{J. Derek Tucker\corref{cor1}}
\ead{dtucker@stat.fsu.edu}
\author[fsu]{Wei Wu}
\ead{wwu@stat.fsu.edu}
\author[fsu]{Anuj Srivastava}
\ead{anuj@stat.fsu.edu}
\cortext[cor1]{Corresponding author. Tel.: +1 850 636 6090; fax: +1 850 235 5374}
\address[nswc]{Naval Surface Warfare Center, Panama City Division - X13, 110 Vernon Avenue, Panama City, FL 32407-7001}
	
\address[fsu]{Department of Statistics, Florida State University, Tallahassee, FL 32306} 

\begin{abstract}
Constructing generative models for functional observations is an important task in statistical functional analysis.
In general, functional data contains both phase (or $x$ or horizontal) 
and amplitude (or $y$ or vertical) variability. 
Traditional methods often ignore the phase variability and 
focus solely on the amplitude variation, using cross-sectional techniques such as fPCA for dimensional reduction and data modeling.  
Ignoring phase variability leads to a loss of structure in the data and inefficiency in data models.  
This paper presents an approach that relies on separating the phase ($x$-axis) and amplitude ($y$-axis), 
then modeling these components using joint distributions. 
This separation, in turn, is performed using a technique called {\it elastic shape analysis of curves}
that involves a new mathematical representation of functional data.   
Then, using individual fPCAs, one each for phase and amplitude components, 
while respecting the nonlinear geometry of the phase representation space; impose 
joint probability models on principal coefficients of these components.
These ideas are demonstrated using random sampling, for models estimated from 
simulated and real datasets, and show their superiority over models that ignore phase-amplitude separation.
Furthermore, the generative models are applied 
to classification of functional data and achieve high performance in applications involving
SONAR signals of underwater objects, handwritten signatures, and periodic body movements recorded by smart phones.
\end{abstract}

\begin{keyword}
	amplitude variability \sep function alignment \sep function principal component analysis \sep functional data analysis \sep generative model \sep phase variability
\end{keyword}
\end{frontmatter}

\section{Introduction} 
The problem of statistical analysis 
and modeling of data in function spaces is important in applications arising 
in nearly every branch of science, including signal processing, geology, biology, and chemistry. 
The observations here are time samples of real-valued functions on an observation interval, and
to perform effective data analysis it is desirable to have a generative, probabilistic model for these observations.  
The model is expected to properly and parsimoniously characterize the nature and variability in the data. 
It should also lead to efficient procedures for conducting hypothesis tests, performing bootstraps, and making forecasts.   
An interesting aspect of functional data is that underlying
variability can be ascribed to two sources. 
In a sample data the given functions may not be perfectly aligned and the mechanism for alignment is 
an important topic of research.
The variability exhibited in functions after alignment is termed the amplitude (or $y$ or vertical) variability and
the warping functions that are used in the alignment are said to capture the phase (or $x$ or horizontal) variability.
A more explicit mathematical definition of amplitude and phase variability will be made in Section~\ref{sec:theory}.
It is imperative that any technique for analysis or modeling of functional data should take 
both these variabilities into account. 

\subsection{Need for Phase-Amplitude Separation}
Many of the current methods for analyzing functional data ignore the phase variability. They 
implicitly assume that the observed functions are already temporally aligned and all the variability is 
restricted only to the $y$-axis.
A prominent example of this situation is 
functional principal component analysis (fPCA) (see e.g., \cite{ramsay-silverman-2005}) 
that is used to discover dominant modes of 
variation in the data and has been extensively used in modeling functional observations.   
If the phase variability is ignored, the resulting model may fail to capture patterns present in the data and will lead to 
inefficient data models.

\begin{figure}[H] 
 \centering
 \subfloat[Original Data]{\includegraphics[height=1.2in]{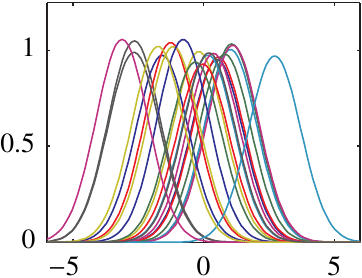}}\hspace{.5em}
 \subfloat[Original Mean]{\includegraphics[height=1.2in]{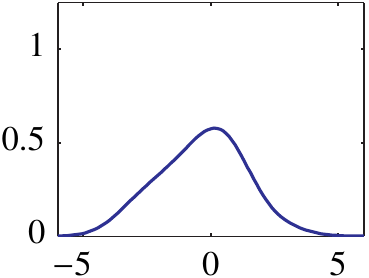}}\hspace{.5em}
 \subfloat[Original Random Samples]{\includegraphics[height=1.2in]{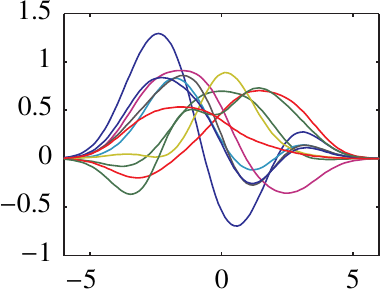}}\\
  \subfloat[Aligned Mean]{\includegraphics[height=1.2in]{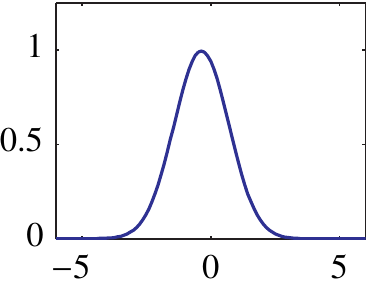}}\hspace{.5em}
  \subfloat[Samples from Proposed Models]{\includegraphics[height=1.2in]{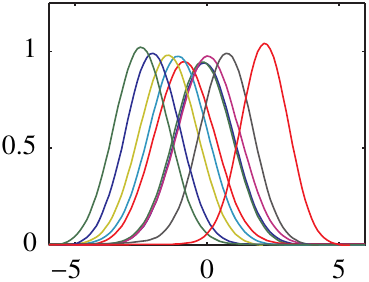}}\\
 \caption{Samples drawn from a Gaussian model fitted to the principal components for the unaligned and aligned data.}
 \label{fig:example}
\end{figure}

Fig.~\ref{fig:example} provides an illustration of this using simulated functional data. This data was
simulated using the equation $y_i(t) = z_i e^{-(t-a_i)^2/2}$,
$t \in [-6, 6], ~i=1,2,\dots, 21$, where $z_{i}$ is {\it i.i.d.} $\mathcal{N}(1, (0.05)^2)$ and $a_i$ is {\it i.i.d.} $\mathcal{N}(0, (1.25)^2)$.
The top-left plot shows the original data; each sample function here is a unimodal function with slight variability in height and a large variability in the 
peak placement.
One can attribute different locations of the peak to the phase variability.
If one takes the cross-sectional mean of this data, ignoring the phase variability,
the result is shown in the top-middle plot. 
The unimodal structure is lost in this mean function with large amount of stretching.
Furthermore, if one performs fPCA on this data and imposes the standard independent normal
models on fPCA coefficients (details of this construction are given later), the resulting model will lack this unimodal structure.
Shown in the top-right plot are random samples generated from such a probability model on the function space
where a Gaussian model is imposed on the fPCA coefficients.
These random samples are not representative of the original data; the essential shape of the function is lost,
with some of the curves having two, three, or even more peaks.

The reason why the underlying unimodal pattern is not retained in the model is that the phase
variability was ignored. We argue that a proper technique is to separate the phase and amplitude
variability, using techniques for functional alignment, and then develop a probability model for
each component. While postponing details for later, we show results obtained by a separation-based approach in the bottom row. 
The mean of the aligned functions is shown in the bottom left panel of Fig. \ref{fig:example}. 
Clearly retained is the unimodal structure and so do the random samples under a framework that model the
phase and amplitude variables individually. 
Some random samples are shown in the bottom right panel, these displays are simply meant to motivate the framework and the mathematical details are
provided later in the paper.
This example clearly motivates the need for function alignment for modeling
functional data that contains phase variability.

\subsection{Past Literature on Phase-Amplitude Separation}
This brings up an important question: How to separate the phase and amplitude components in 
a given dataset? While this is a topic of ongoing research, a number of techniques have already been 
discussed in the literature. The main difference between them lies in the choice of the cost 
function used in the alignment process. The different cost functions suggested in the statistics literature 
including area-under-the-curve matching (\cite{muller-JASA:2004,muller-biometrika:2008}),
minimum second-eigenvalue (MSE) (\cite{kneip-ramsay:2008}), moment based matching (MBM) (\cite{james:10}), 
self-modeling registration (\cite{gervini-gasser-RSSB:04}) and  
$k$-means alignment and clustering (\cite{sangalli-etal:2010,sangalli-etal2:2010}).

In the meantime several other communities, often outside statistics,  have studied 
registration of functions in one or higher dimensions, e.g., in matching MRI images (\cite{Christensen:2001,Tagare:2009,Beg:2005}),  
shape analysis of curves
(\cite{klassen-srivastava-etal:04,michor-mumford-shape,joshi-klassen-cvpr:07,srivastava-kalseen-joshi-jermyn:11,kurtek-et-all:12,art:younes2,art:younes}), 
shape analysis of surfaces (\cite{kurtek:2010}), etc. 
The problem of curve registration is especially relevant for phase-amplitude
separation needed in functional data analysis since the case for $\real^1$ is essentially that for real valued functions!
We will adapt a shape-analysis approach that has been termed {\it elastic shape analysis} 
(\cite{joshi-klassen-cvpr:07,srivastava-kalseen-joshi-jermyn:11,kurtek-et-all:12}).
Although these methods have been developed for alignment of curves in $\real^n$, 
their application to functional data analysis has been explained in \cite{kurtek-wu-srivastava-NIPS:2011,kaziska-FANOVA:10,srivastava-etal-JASA:2011}.
The basic idea in this method is to introduce a mathematical representation, called the {\it square-root slope function} or SRSF
(details in the next section) that improves functional alignment and provides fundamental mathematical equalities
that leads to a formal development of this topic. 

The theoretical superiority of  the elastic method comes from the following fact: the alignment of functions is based on a cost term that is a proper distance. 
Thus satisfying all desired properties in alignment, 
such as symmetry (optimal alignment of $f$ to $g$ is same as that of $g$ to $f$), positive definiteness 
(the cost term between any two functions is nonnegative and it equals zero if and only if one can be perfectly aligned to the other), 
and the triangle inequality. None of the current methods in the 
statistics literature (e.g., \cite{kneip-ramsay:2008,james:10,muller-JASA:2004,gervini-gasser-RSSB:04,muller-biometrika:2008})
have these properties. 
In fact, most of them are not even symmetric in their alignment. Additionally, many past methods perform component separation and 
fPCA in two distinct steps, under different metrics, while in elastic shape analysis it is performed jointly under
the same metric. 
In addition to these theoretical advantages, 
we have also emphasized the experimental superiority of elastic curve analysis using a large number of datasets 
in this paper. 

Another important issue, encountered in modeling phase variability, is to characterize the geometry of the 
phase space. Generally speaking, phase variability is represented by a warping function 
$\gamma$ that satisfies certain properties such as boundary conditions, invertibility, smoothness, 
and smoothness of its inverse. 
\cite{ramsay-silverman-2005} represent $\gamma$ using a basis set in the log-derivative space, i.e., 
$\log(\dot{\gamma}(t)) = \sum_{i} \alpha_i b_i(t)$. Some others force $\gamma$ to be a piecewise
linear function with positive derivatives (\cite{muller-JASA:2004}) and even linear functions (\cite{sangalli-etal:2010,sangalli-etal2:2010}).  
It becomes clear that boundary conditions, combined with the smoothness and invertibility 
requirements, restrict the set of allowable warping functions to a nonlinear space. 
Although it seems natural, the use of nonlinear geometry of this set in establishing a 
metric for comparing warping functions and for performing fPCA has seldom been 
studied in the functional data analysis literature. \cite{srivastava-kalseen-joshi-jermyn:11}
have studied a square-root derivative representation, similar to the one suggested
by \cite{bhattacharya-43}, for converting this set into a sphere and 
analyzing warping functions as elements of a Hilbert sphere. The paper 
\cite{srivastava-etal-Fisher-Rao-CVPR:2007} demonstrates the advantages of using 
square-root derivative over the log-derivative representation of warping functions. 

\subsection{Proposed Framework}
After the separation of phase and amplitude components, we will define two types of distances.
One is a $y$-distance, defined to measure amplitude differences between any two functions (and independent 
of their phase variability) and computed
as the $\ltwo$ distance between the SRSFs of the aligned functions.   
The other is an $x$-distance, or the distance on their phase components, 
that measures the amount of warping needed to align the functions.
We will show that either of these distances provides useful measures for 
computing summary statistics, for performing fPCA, and for discriminating 
between function classes.

The main contribution of this paper is a 
modeling framework to characterize 
functional data using phase and amplitude separation.  
The basic steps in this procedure are: 1) Align the original functional data and obtain the aligned functions (describing amplitude variability) and the 
warping functions (describing phase variability).  2) Estimate 
the sample means and covariance on the phase and amplitude, respectively.  
This step uses a nonlinear transformation on the data to enable
use of $\ltwo$ norm (and cross-sectional computations) for generating summary statistics (see Section~\ref{sec:pca}); 
3) Based on the estimated summary statistics, perform fPCA on the phase and amplitude, respectively;
4) Model the original data by using joint Gaussian or non-parametric models on both fPCA representations; 
5) As a direct application, the model can be used to perform likelihood-based classification of  functional data.  
We will illustrate this application using several data sets which include a simulated data set, a signature data set from \cite{Yeung-et-al:2004}, an iPhone action data set from \cite{mccall-reddy-shah}, and a SONAR data set.

The rest of this paper is organized as follows: Section~\ref{sec:theory} presents the differential geometric approach for phase and 
amplitude separation adapted from  \cite{srivastava-kalseen-joshi-jermyn:11,joshi-klassen-cvpr:07} and 
explained in \cite{srivastava-etal-JASA:2011,kurtek-wu-srivastava-NIPS:2011}. 
Section~\ref{sec:pca} presents the functional principal component analysis 
of these phase and amplitude components, and statistical modeling of their principal coefficients.  
These modeling results are presented in Section~\ref{sec:modelingresult}. 
Section~\ref{sec:classification} describes classification of functional data
using the developed models on real data sets, and compares results with some conventional methods. 
Finally, conclusions and observations are offered in Section~\ref{sec:conclusion}. 
We have developed and R package \verb+fdasrvf+ implementing the proposed functional alignment and fPCA method \cite{fdasrvf}; this package is available on the CRAN archive.

\section{Phase and Amplitude Separation Using Elastic Analysis} \label{sec:theory} 
In this section, we adapt a method introduced for elastic 
shape analysis of curves  in \cite{srivastava-kalseen-joshi-jermyn:11,joshi-klassen-cvpr:07} 
to the problem of functional data  alignment. The details are 
presented in companion papers  \cite{srivastava-etal-JASA:2011,kurtek-wu-srivastava-NIPS:2011}.
This comprehensive framework addresses three important goals: (1) completely automated alignment of functions using nonlinear 
time warping, (2) separation of phase and amplitude components of functional data, and 
(3) derivation of individual phase and amplitude metrics for comparing and classifying functions. 
For a more comprehensive introduction to this theory, including asymptotic results and estimator convergences, we refer the reader to these two papers as we will only present the algorithm here.

\subsection{Mathematical Representation of Functions} 
Let $f$ be a real-valued function with the domain $[0,1]$; the domain can easily be transformed to any other interval. 
For concreteness, only functions that are absolutely continuous on $[0,1]$ will be considered; let $\F$ denote the set of all such functions. 
In practice, since the observed data are discrete, this assumption is not a restriction. 
Also, let $\Gamma$ be the set of boundary-preserving diffeomorphisms of the unit interval $[0,1]$: $\Gamma = \{\gamma: [0,1] \to [0,1] |~\gamma(0) = 0,~ \gamma(1)=1,\gamma~\textnormal{is a diffeomorphism} \}$. 
Elements of $\Gamma$ play the role of warping functions. For any $f \in \F$ and $\gamma \in \Gamma$, the 
composition $f \circ \gamma$ denotes the time-warping of $f$ by $\gamma$. With the composition operation, the 
set $\Gamma$ is a group with the identity element $\gamma_{id}(t) = t$. This is an important observation 
since the group structure of $\Gamma$ is seldom utilized in past papers on functional data analysis.  

In a pairwise alignment problem, the goal is to align any two functions $f_1$ and $f_2$. 
A majority of past methods use cost terms of the type $(\inf_{\gamma \in \Gamma}
\| f_1 - f_2 \circ \gamma\|)$ to perform this alignment. Here $\| \cdot \|$ denotes
the standard $\ltwo$ norm. However, this alignment is neither symmetric nor 
positive definite. To address this and other related problems, 
\cite{srivastava-kalseen-joshi-jermyn:11} introduced  a mathematical expression for representing a function. 
This function, $q: [0,1] \to \real$, is called the {\it square-root slope function} or SRSF of $f$, and is defined in the following form:  
\[q(t) = \mbox{sign}(\dot{f}(t)) \sqrt{ |\dot{f}(t)|}\ .\]
It can be shown that if the function $f$ is absolutely continuous, then the resulting SRSF is square-integrable 
(see \cite{robinson-2012} for a proof). 
Thus, we will define $\ltwo([0,1],\real)$, or simply $\ltwo$, to be the set of all SRSFs. 
For every $q \in \ltwo$ and a fixed $t \in [0, 1]$, the function $f$ can be obtained precisely using the equation:
$	f(t) = f(0) + \int_{0}^{t} q(s) |q(s)| ds$, since $q(s) |q(s)| = \dot{f}(s)$.
Therefore, the mapping from ${\cal F}$ to $\ltwo \times \real$, 
given by $f  \mapsto (q,f(0))$ is a bijection (see \cite{robinson-2012}). 
The most important property of this framework is the following. 
If we warp a function $f$ by $\gamma$, the SRSF of $f \circ \gamma$ is given by: $\tilde{q}(t) = (q,\gamma)(t) = q(\gamma(t))\sqrt{\dot{\gamma}(t)}$. 
With this expression it can be shown that for any $f_1, f_2 \in \F$ and $\gamma \in \Gamma$, 
we have
\begin{equation}
\| q_1 - q_2 \| = \| (q_1, \gamma) - (q_2, \gamma) \|\ , \label{eq:isometry}
\end{equation}
where $q_1, q_2$ are SRSFs of $f_1, f_2$, respectively. This is called the {\it isometry} property
and it is central in suggesting a new cost term for pairwise registration of functions:
$\inf_{\gamma \in \Gamma}
\| q_1 - (q_2, \gamma)\|$. 
This equation suggests we can register (or align) the SRSFs of any two functions first and 
then map them back to $\F$ to obtain registered functions.  
The advantage of this choice is that it is symmetric, 
positive definite, and satisfies the triangle inequality. 
Technically, it forms a proper distance\footnote{We note that restriction of $\ltwo$ metric to 
SRSFs of functions whose first derivative 
is strictly positive, e.g., cumulative distribution functions, is exactly the classical Fisher-Rao Riemannian metric 
used extensively in the statistics community \cite{rao:45,Cencov82,Kass97,efron-annals:75,amari85}.} on the quotient 
space $\ltwo/\Gamma$. 
We mention that this cost function has a built-in regularization term and does 
not require any additional penalty term. 
Please refer to papers \cite{srivastava-etal-JASA:2011,kurtek-wu-srivastava-NIPS:2011} for more details.
In case one wants to control the amount of warping or {\it elasticity} this can be done 
as described in \cite{wu-srivastava:2011}.

The isometric property in Eqn. \ref{eq:isometry} leads to a distance between functions that is {\it invariant} to their random warpings:
\begin{definition}
	[Amplitude or $y$-distance] For any two functions $f_1,\ f_2 \in \F$ and the 
	corresponding SRSFs, $q_1, q_2 \in \ltwo$, we define the amplitude or the $y$-distance $D_y$ to be: 
	\[D_y(f_1, f_2) = \inf_{\gamma \in \Gamma} \|q_1 - (q_2 \circ \gamma)\sqrt{\dot{\gamma}}) \|.\]
\end{definition}
It can be shown that for any $\gamma_1, \gamma_2 \in \Gamma$, we have 
$D_y(f_1 \circ \gamma_1, f_2 \circ \gamma_2) = D_y(f_1, f_2)$. 

\noindent {\bf Optimization Over $\Gamma$}: 
The minimization over $\Gamma$ can be performed in many ways. In case $\Gamma$ is represented 
by a parametric family, then one can use the parameter space to perform the estimation as \cite{kneip-ramsay:2008}.
However, since 
$\Gamma$ is a nonlinear manifold, it is impossible to express it completely in a parametric vector space. 
In this paper we use the standard Dynamic Programming algorithm (\cite{bertsekas-DP}) to solve for an optimal $\gamma$. 
It should be noted that for any fixed partition of the interval $[0,1]$, this algorithm provides the 
exact optimal $\gamma$ that is restricted to the graph of this partition. 

\subsection{Karcher Mean and Function Alignment} \label{sec:karcher}
In order to separate phase and amplitude variability in functional data, we need a notion of the mean of functions.
Basically, first we compute a mean function and in the process warp the given functions to match the mean function.
Since we have a {\bf proper} distance in $D_y$, in the sense that it is invariant to random warping, we can 
use that to define this mean. 
For a given collection of functions $f_1, f_2, \dots, f_n$, let $q_1, q_2, \dots, q_n$ denote their SRSFs, respectively. 
Define the Karcher mean of the given function as a local minimum of the following cost functions: 
\begin{eqnarray}
		\mu_f &=& \argmin_{f \in {\cal F}} \sum_{i=1}^n D_y(f, f_i)^2\ \ \\		 
		\mu_q &=&  \argmin_{q \in \ltwo} \sum_{i=1}^n \left( \inf_{\gamma_i \in \Gamma} \| q - (q_i, \gamma_i) \|^2 \right)\ . \label{eq:Karcher-min} 
\end{eqnarray}
(This Karcher mean has also been called by other names such as the Frechet mean, intrinsic mean or the centroid.) 
These are two equivalent formulations, one in the function space $\F$ and other in 
the SRSF space $\ltwo$, i.e., $\mu_q = \mbox{sign}(\dot{\mu}_f) \sqrt{ |\dot{\mu}_f|}$. 
Note that if ${\mu}_f$ is a minimizer of the cost function, then so is ${\mu}_f \circ \gamma$ for 
any $\gamma \in \Gamma$ since $D_y$ is invariant to random warpings of its input variables. 
So, we have an extra degree of freedom in choosing an arbitrary element of the set $\{ \mu_f \circ \gamma | \gamma \in \Gamma\}$. 
To make this choice unique, we can define a special element of this class as follows. 
Let $\{\gamma_i^*\}$ denote the set of optimal warping functions, one for each $i$, in Eqn. \ref{eq:Karcher-min}.
Then, we can choose the $\mu_f$ to that element of its class 
such that the mean of $\{\gamma_i^*\}$, denoted by
$\gamma_{\mu}$,  is $\gamma_{id}$, the identity element.
(The notion of the mean of warping functions and its computation are described later in Section \ref{sec:x-var} and summarized in Algorithm 2).
The algorithm for computing the Karcher mean ${\mu}_f$ of SRSFs is given in Algorithm 1, where the iterative update in Steps 2-4 is based on the gradient of the cost function given in Eqn. \ref{eq:Karcher-min}.

\begin{algorithm}[htbp]
	{\bf Algorithm 1: Phase-Amplitude Separation} 
	\label{algo:mean-svrf} 
	\begin{enumerate}
		\item Compute SRSFs $q_1, q_2, \dots, q_n$ of the given $f_1, f_2, \dots, f_n$ 
		 and select $\mu = q_i$, where $i =\argmin_{1\le i \le n} || q_i - \frac{1}{n} \sum_{j=1}^n q_j||$. 
		\item For each $q_i$ find the $\gamma_i^*$ such that $\gamma_i^* = \argmin_{\gamma \in \Gamma} \left( \| \mu - (q_i \circ \gamma) \sqrt{\dot{\gamma}}\| \right)$. The solution to this optimization comes from the dynamic programming algorithm.
		\item Compute the aligned SRSFs using $\tilde{q}_i \mapsto (q_i \circ \gamma_i^*) \sqrt{\dot{\gamma_i^*}}$. 
		\item If the increment $\| \frac{1}{n} \sum_{i=1}^n \tilde{q}_i - \mu\|$ is small, then continue. Else, update the mean using $\mu \mapsto \frac{1}{n} \sum_{i=1}^n \tilde{q}_i$ and return to step 2.
		\item  The function $\mu$ represents a whole equivalence class of solutions and now we
		select the preferred element ${\mu}_q$ of that orbit:

		\begin{enumerate}
			\item Compute the mean $\gamma_\mu$ of all $\{\gamma_i^*\}$ (using Algorithm 2). 
			Then compute ${\mu}_q =  (\mu \circ \gamma_{\mu}^{-1})\sqrt{\dot{\gamma_{\mu}^{-1}}}$. \\
			\item Update $\gamma_i^*  \mapsto  \gamma_i^* \circ \gamma_{\mu}^{-1}$.  
			Then compute the aligned SRSFs using $\tilde{q}_i \mapsto (q_i \circ \gamma_i^*) \sqrt{\dot{\gamma_i^*}}$. 
		\end{enumerate}
	\end{enumerate}
\end{algorithm}
This procedure results in three items: 
\begin{enumerate}
	\item $\mu_q$, preferred element of the Karcher mean class $\{({\mu}_q, \gamma) | \gamma \in \Gamma\}$, 
	\item $\{ \tilde{q}_i\}$, the set of aligned SRSFs, and
	\item $\{ \gamma_i^*\}$, the set of optimal warping functions.
\end{enumerate} 
From the aligned SRSFs, one can compute individual aligned functions using: $\tilde{f}_i(t) = f_i(0) + \int_{0}^t \tilde{q}_i(s) |\tilde{q}_i(s)|\, ds$.

\begin{figure}[htbp]
	\centering
  \subfloat[Original SRSFs $\{q_i\}$]{\includegraphics{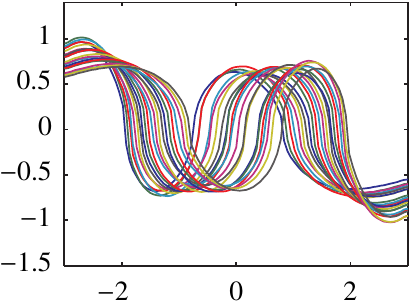}\label{fig:simu_q}}\hspace{.5em}
  \subfloat[Warped Data SRSFs $\{ \tilde{q}_i\}$]{\includegraphics{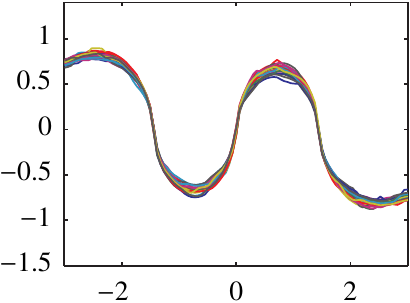}\label{fig:simu_qn}}\hspace{.5em}
  \subfloat[Warping Functions $\{ \gamma_i^*\}$]{\includegraphics{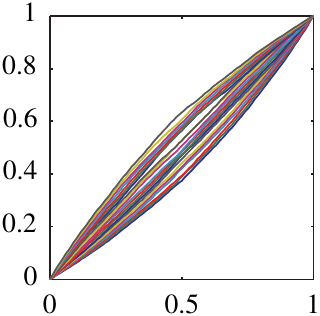}\label{fig:simu_gam}}\\
	\subfloat[Original Data $\{f_i\}$]{\includegraphics{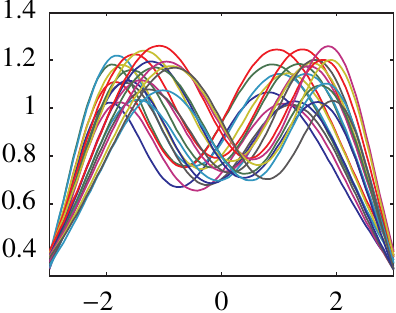}\label{fig:simu_f}}\hspace{.5em}
	\subfloat[Warped Data $\{ \tilde{f}_i\}$]{\includegraphics{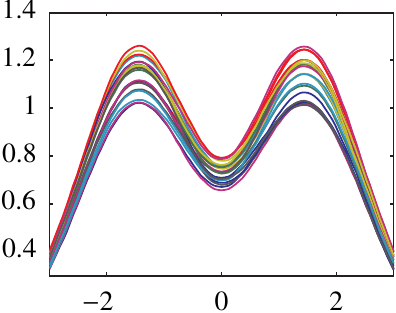}\label{fig:simu_fn}}
	\caption{Alignment of the simulated data set using Algorithm 1.}
	\label{fig:mean-result-sim}
\end{figure}

To illustrate this method we run the algorithm on the data previously used in \cite{kneip-ramsay:2008}. 
The individual functions are given by: $y_i(t) = z_{i,1} e^{-(t-1.5)^2/2} + z_{i,2}e^{-(t+1.5)^2/2}$, $t \in [-3, 3], ~i=1,2,\dots, 21$, where $z_{i,1}$ and $z_{i,2}$ are {\it i.i.d.} $\mathcal{N}(1, (0.25)^2)$. 
(Note that although the elastic framework was developed for functions on $[0,1]$, it can easily be adapted to an arbitrary interval). 
Each of these functions is then warped according to: $\gamma_i(t) = 6\left(\frac{e^{a_i(t+3)/6} -1}{e^{a_i} - 1}\right) - 3$ if  $a_i \neq 0$, otherwise $\gamma_i = \gamma_{id}$ ($\gamma_{id}(t) = t$ is the identity warping). 
Here $a_i$ are equally spaced between $-1$ and $1$, and the observed functions are computed using $x_i(t) = y_i \circ \gamma_i (t)$. 
A set of 21 such functions forms the original data and is shown in Panel \subref{fig:simu_f} of Fig.~\ref{fig:mean-result-sim} with corresponding SRSFs in Panel \subref{fig:simu_q}. 
Panel \subref{fig:simu_qn} presents the resulting aligned SRSFs using our method $\{ \tilde{q}_i\}$ and Panel \subref{fig:simu_gam} plots the corresponding warping functions $\{ \gamma_i^*\}$.
The corresponding aligned functions $\{ \tilde{f}_i\}$ is shown in Panel \subref{fig:simu_fn}. 
It is apparent that the plot of $\{\tilde{f}_i\}$ shows a tighter alignment of functions with sharper peaks and valleys, and thinner
band around the mean. 
This indicates that the effects of warping generated by the $\gamma_i$s have been completely removed and only the randomness from the $y_i$s remain.

\begin{figure}[t]
\begin{center}
	\subfloat[$\{f_i\}$]{\includegraphics{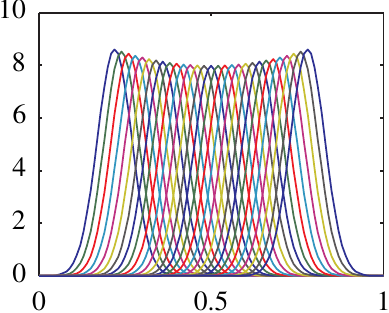}}\hspace{.25em}
	\subfloat[Elastic $\{ \tilde{f}_i\}$]{\includegraphics{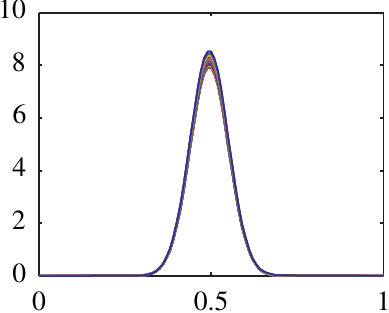}}\hspace{.25em}
	\subfloat[MBM $\{ \tilde{f}_i\}$]{\includegraphics{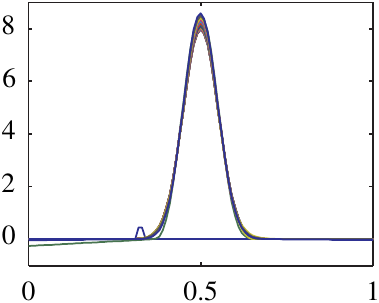}}\hspace{.25em}
	\subfloat[MSE $\{ \tilde{f}_i\}$]{\includegraphics{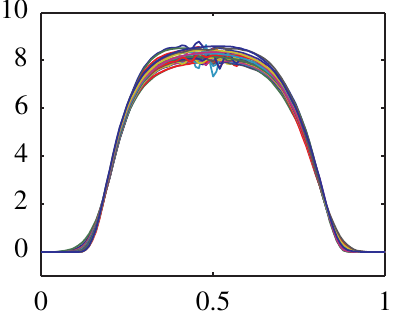}}\\
  \subfloat[$\{f_i\}$]{\includegraphics{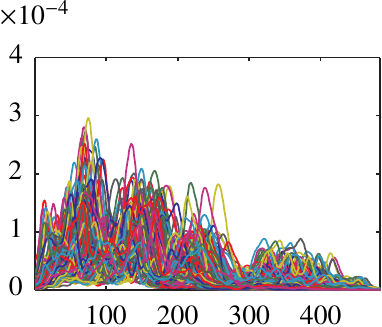}}\hspace{.25em}
  \subfloat[Elastic $\{ \tilde{f}_i\}$]{\includegraphics{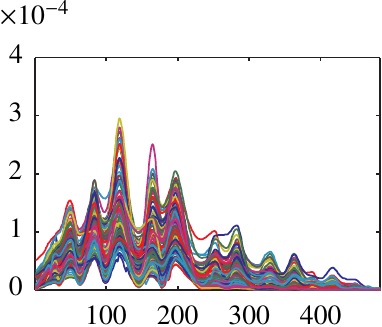}}\hspace{.25em}
  \subfloat[MBM $\{ \tilde{f}_i\}$]{\includegraphics{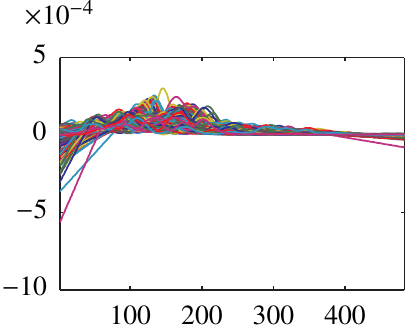}}\hspace{.25em}
  \subfloat[MSE $\{ \tilde{f}_i\}$]{\includegraphics{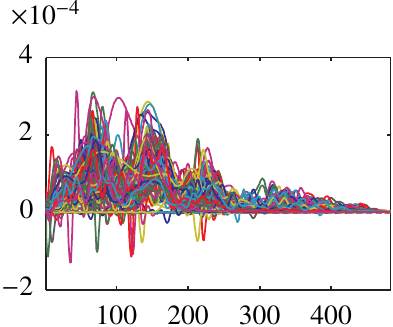}}
	\caption{Comparison of alignment algorithms on a difficult unimodal data set (top row) and a real SONAR data set (bottom row).}
	\label{fig:alignresults}
\end{center}
\end{figure}

We also compare the performance of Algorithm 1 with some published methods including; the MBM method of \cite{james:10} and the MSE method of 
\cite{ramsay-silverman-2005} on a more difficult simulated data and a real SONAR data set.
The original simulated data are shown in Fig.~\ref{fig:alignresults}(a) and the data consists of 39 unimodal functions which have been warped with equally-spaced centers along the $x$-axis and have slight variation in peak-heights along the $y$-axis. 
Fig.~\ref{fig:alignresults}(b)-(d) present the alignment results for our elastic method, the MBM method, and the MSE method, respectively. 
The original SONAR data are shown in Fig.~\ref{fig:alignresults}(e) and the data consists of 131 measured SONAR signals that contain measurement ambiguity. 
Fig.~\ref{fig:alignresults}(f)-(h) present the alignment results for our elastic method, the MBM method, and the MSE method, respectively. 
For the simulated data the elastic method performs the best while the MBM method performs fairly well with a little higher standard deviation.
The MBM method and the MSE method both have a few numerical issues that lead to blips in the functions.
For the SONAR data only the elastic method performs well, as MBM and MSE methods fail to align the data at all.
We can also quantify the alignment  performance using the decrease in the cumulative cross-sectional variance of the aligned functions.
For any functional dataset $\{ g_i(t),i=1,2,\dots,n, t \in [0,1]\}$, let 
\[\mbox{Var}(\{g_i\}) = \frac{1}{n-1} \int_0^1 \sum_{i=1}^n \left(g_i(t) - \frac{1}{n}\sum_{i=1}^n g_i(t) \right)^2 dt\ ,\]
denote the cumulative cross-sectional variance in the given data. With this notation, we define: 
\[\mbox{Original Variance} = \mbox{Var}(\{f_i\}),
\ \ \ \mbox{Amplitude Variance} = \mbox{Var}(\{\tilde{f}_i\}),\ \mbox{Phase Variance} = \mbox{Var}(\{ \mu_f \circ \gamma_i \})
\ .\]
The phase- and amplitude-variances for the different alignment algorithms shown in Fig.~\ref{fig:alignresults} is listed below in
Table \ref{tab:align_variance} with the simulated unimodal data on the top two rows and the SONAR data on the bottom two rows:
\begin{table}[htbp]
	\begin{center}
	\begin{tabular}{c c c c c c}
	\hline
		Data & Component &Original Variance & Elastic Method & MBM  & MSE  \\ \hline \hline
		 Unimodal & Amplitude-variance & 4.33 & {\bf 0.004} & 0.23 & .02 \\ 
		\cline{2-6}
		 & Phase-variance & 0& 4.65 & 4.31 & 4.54 \\
		\hline \hline
    SONAR Data & Amplitude-variance & 2.89e-5 & {\bf 1.53e-5} & 3.02e-5 & 2.42e-5 \\ 
    \cline{2-6}
    & Phase-variance & 0& 1.48e-5 & 1.30e-5 & 1.36e-5 \\
    \hline
    	\end{tabular}
	\end{center}
	\caption{The comparison of the amplitude variance and phase variance for different alignment algorithms on the Unimodal and SONAR data set.}
	\label{tab:align_variance}
	\end{table}	
Based on its superior performance and theoretical advantages, we choose the elastic method for 
separating the phase and amplitude components. For additional experiments and asymptotic analysis of 
this method, please refer to  \cite{srivastava-etal-JASA:2011,kurtek-wu-srivastava-NIPS:2011}. 

\section{Analysis and Modeling of Components} \label{sec:pca} 
Having separated functional data into phase and amplitude components, we focus on the 
task of developing their generative models.  

\subsection{Phase-Variability: Analysis of Warping Functions} \label{sec:x-var} 
First, we would like to study the phase-variability of the given functions, available to us in the form of 
the warping functions $\{ \gamma_i^*\}$ resulting from Algorithm 1. 
An explicit statistical modeling of the warping functions can be of interest to an analyst since they represent the phase-variability of the original data. 
As mentioned earlier, the space of warping functions, $\Gamma$, is a nonlinear manifold and cannot 
be treated as a Hilbert space directly. 
Therefore, we will use tools from differential geometry to be able to perform statistical analysis and modeling of the warping functions. 
This framework has been used previously but in different application areas, 
e.g., modeling parameterizations of curves \cite{srivastava-jermyn-PAMI:09} and studies of execution rates of human activities in 
videos \cite{ashok-srivastava-etal-TIP:09}. It also relates to the square-root representation of probability densities
introduced by \cite{bhattacharya-43}. 

\begin{figure}[t]
\begin{center}
  \includegraphics{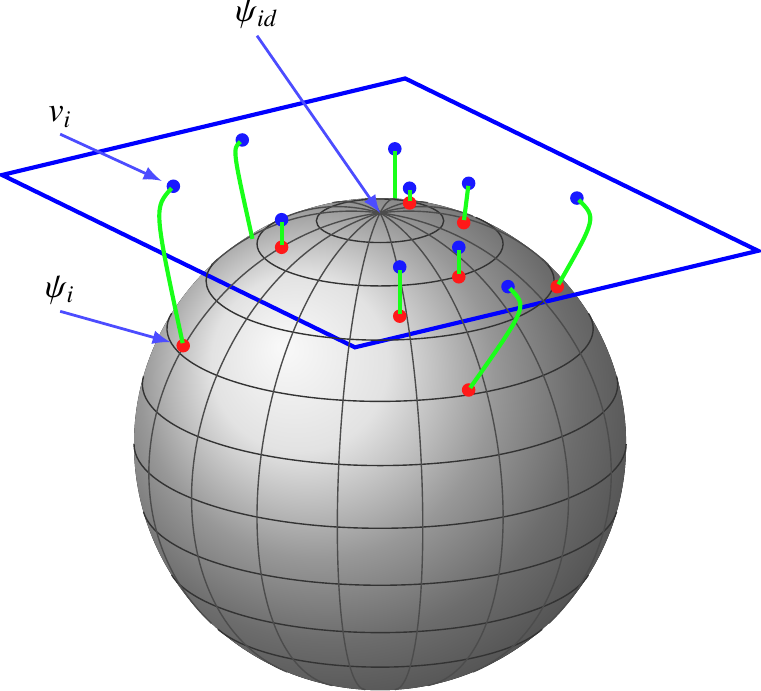}
  \caption{Depiction of the SRSF space of warping functions as a sphere and a tangent space at identity $\psi_{id}$.} 
  \label{fig:sphere-map}
\end{center}
\end{figure}

Let $\gamma_1, \gamma_2, \dots, \gamma_n \in \Gamma$ be a set of observed warping functions. 
Our goal is to develop a probability model on $\Gamma$ that can be estimated from the data directly. 
There are two problems in doing this in a standard way: (1) $\Gamma$ is a nonlinear manifold, and (2) it is infinite dimensional. 
The issue of nonlinearity is handled using a convenient transformation which coincidentally is similar to the definition of SRSF, and the issue of infinite dimensionality is handled using dimension reduction, 
e.g., fPCA, which we will call {\it horizontal} fPCA. 
We are going to represent an element $\gamma \in \Gamma$ by the square-root of its derivative $\psi = \sqrt{\dot{\gamma}}$. 
Note that this is the same as the SRSF defined earlier for $f_i$s and takes this form since $\dot{\gamma} > 0$. 
The identity map $\gamma_{id}$ maps to a constant function with value $\psi_{id}(t) = 1$. Since $\gamma(0) = 0$, the mapping from $\gamma$ to $\psi$ is a bijection and one can reconstruct $\gamma$ from $\psi$ using $\gamma(t) = \int_0^t \psi(s)^2 ds$. 
An important advantage of this transformation is that since $\| \psi\|^2 = \int_0^1 \psi(t)^2 dt = \int_0^1 \dot{\gamma}(t) dt = \gamma(1) - \gamma(0) = 1$, the set of all such $\psi$s is a Hilbert sphere $\s_{\infty}$, a unit sphere in the Hilbert space $\ltwo$. 
In other words, the square-root representation simplifies the complicated geometry of $\Gamma$ to a unit sphere.  
The distance between any two warping functions is exactly the arc-length between their corresponding SRSFs on the unit 
sphere $\s_{\infty}$: 
$$
D_{x}(\gamma_1, \gamma_2) = d_{\psi}(\psi_1, \psi_2) \equiv \cos^{-1}\left(\int_0^1 \psi_1(t) \psi_2(t) dt \right)\ .
$$ 
Fig.~\ref{fig:sphere-map} shows an illustration 
of the SRSF space of warping functions as a unit sphere. 

The definition of a distance on $\s_{\infty}$ helps define a Karcher mean of sample points on $\s_{\infty}$. 
\begin{definition}
	For a given set of points $\psi_1, \psi_2, \dots, \psi_n \in \s_{\infty}$, their Karcher mean in 
	$\s_{\infty}$ is defined to be a local minimum of the cost function $\psi \mapsto \sum_{i=1}^n d_{\psi}(\psi, \psi_i)^2$.
\end{definition}
Now we can define the Karcher mean of a set of warping functions using the Karcher mean in $\s_{\infty}$. 
For a given set of warping functions $\gamma_1, \gamma_2, \dots, \gamma_n \in \Gamma$, their Karcher mean
in $\Gamma$ is $\bar{\gamma}(t) \equiv \int_0^t \mu_{\psi}(s)^2 ds$ where $\mu_{\psi}$ is the Karcher mean of $\sqrt{\dot \gamma_1}$, $\sqrt{\dot \gamma_2}$, $\dots$, $\sqrt{\dot \gamma_n}$ in $\s_{\infty}$. 
The search for this minimum is performed using Algorithm 2: 
\begin{algorithm}[htbp]
	\label{algo:x-mean} 
	{\bf Algorithm 2: Karcher Mean of Warping Functions} \\
Let $\psi_i = \sqrt{\dot{\gamma}_i}$ be the SRSFs for the given warping functions. Initialize $\mu_{\psi}$ to be one of the $\psi_i$s or simply $w/\|w\|$, where $w = \frac{1}{n}\sum_{i=1}^n \psi_i$. 
	\begin{enumerate}
		\item For $i=1,2,\dots,n$, compute the shooting vector $v_i = \frac{\theta_i}{\sin(\theta_i)}(\psi_i - \cos(\theta_i) \mu_{\psi})$, $\theta_i = \cos^{-1}\left(\inner{\mu_{\psi}}{\psi}\right)$. By definition, each of these $v_i \in T_{\mu_{\psi}}(\s_{\infty})$.
		\item Compute the average $\bar{v} = \frac{1}{n} \sum_{i=1}^n v_i \in T_{\mu_{\psi}}(\s_{\infty})$. 
		\item If $\|\bar{v}\|$ is small, then continue. Else, update $\mu_{\psi} \mapsto \cos(\epsilon \| \bar{v}\|)\mu_{\psi} + \sin(\epsilon \|\bar{v}\|)\frac{ \bar{v}}{\|\bar{v}\|}$, for a small step size $\epsilon> 0$ and return to Step 1.
		\item Compute the mean warping function using $\bar{\gamma}(t) = \int_0^t \mu_\psi(s)^2 ds$. Stop. 
	\end{enumerate}
\end{algorithm}

Since $\s_{\infty}$ is a nonlinear space (a sphere), one cannot perform principal component analysis on it directly. 
Instead, we choose a vector space tangent to the space, at a certain fixed point, for analysis. 
The tangent space at any point $\psi \in \s_{\infty}$ is given by: $T_{\psi}(\s_{\infty}) = \{v \in \ltwo| \int_0^1 v(t) \psi(t) dt = 0\}$. 
In the following, we will use the tangent space at $\mu_{\psi}$ to perform analysis. 
Note that the outcomes of Algorithm 2 include the Karcher mean $\mu_{\psi}$ and the tangent vectors $\{v_i\} \in T_{\mu_{\psi}}(\s_{\infty})$. 
These tangent vectors, also called the {\it shooting vectors}, are the mappings of $\psi_i$s into 
the tangent space  $T_{\mu_{\psi}}(\s_{\infty})$, as depicted in Fig.~\ref{fig:sphere-map}. 
In this tangent space we can define a sample covariance function: $(t_1, t_2) \mapsto \frac{1}{n-1} \sum_{i=1}^n v_i(t_1) v_i(t_2)$. 
In practice, this covariance is computed using a finite number of points, say $T$, on these functions and one obtains a $T \times T$ sample covariance matrix instead, denoted by $K_{\psi}$. 
The singular value decomposition (SVD) of $K_{\psi} = U_{\psi} \Sigma_{\psi} V_{\psi}^\T$ provides the estimated principal components of $\{ \psi_i\}$: the principal directions $U_{\psi,j}$ and the observed principal coefficients $\inner{v_i}{U_{\psi,j}}$. 
This analysis on $\s_{\infty}$ is similar to the ideas presented in \cite{srivastava-joshi-etal:05} 
although one can also use the idea of principal nested sphere for this analysis \cite{jung-dryden-marron:2012}.

\begin{figure}[htbp]
\begin{center}\mbox{
	 \subfloat[]{\includegraphics{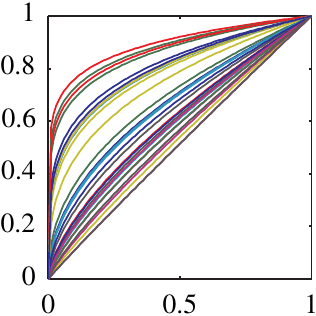}}\hspace{.1em}
		\subfloat[]{\includegraphics{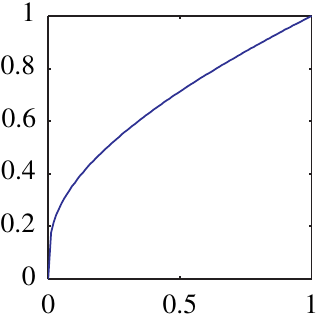}}\hspace{.1em}
		\subfloat[]{\includegraphics{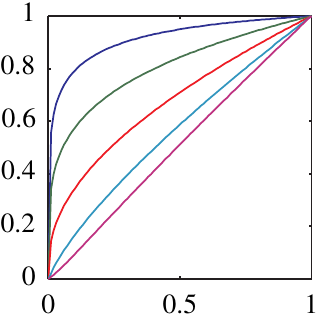}}\hspace{.1em}
		\subfloat[]{\includegraphics{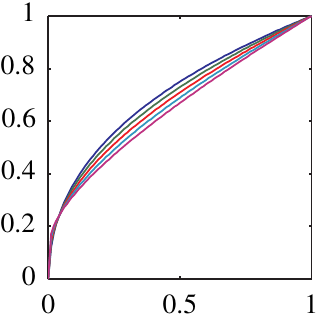}}\hspace{.1em}
		\subfloat[]{\includegraphics{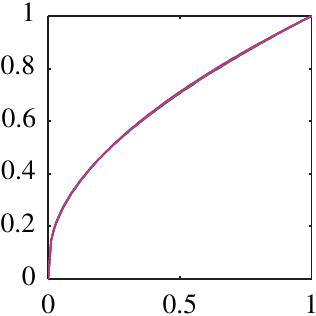}}}
	\caption{From left to right: a) the observed warping functions, b) their Karcher mean, c) the first principal direction, d) second principal direction, and e) third principal direction of the observed data.}
	\label{fig:mean-warping}
\end{center}
\end{figure}

As an example, we compute the Karcher mean of a set of random warping functions. 
These warping functions are shown in the left panel of Fig.~\ref{fig:mean-warping} and their Karcher mean is shown in the second panel.
Using the $\{v_i\}$'s that result from Algorithm 2, we form their covariance matrix $K_{\psi}$ and take its SVD. 
The first three columns of $U_{\psi}$ are used to visualize the principal geodesic paths in the third, fourth, and fifth panels.

\subsection{Amplitude Variability: Analysis of Aligned Functions} 
Once the given observed SRSFs have been aligned using Algorithm 1, they can be statistically analyzed in a standard way (in $\ltwo$)
using cross-sectional computations in the SRSF space. 
This is based on the fact that $D_y$ is the $\ltwo$ distance between the aligned SRSFs. 
For example, one can compute their principal components for the purpose of dimension reduction and statistical modeling using fPCA. 
Since we are focused on the amplitude-variability in this section, we will call this analysis {\it vertical fPCA}.

Let $f_1,\cdots,f_n$ be a given set of functions, and $q_1,\cdots,q_n$ be the corresponding SRSFs, ${\mu}_q$ be their Karcher Mean, 
and let $\tilde{q}_i$s be the corresponding aligned SRSFs using Algorithm 1. 
In performing  vertical fPCA, one should not forget about the variability associated with the initial 
values, i.e., $\{f_i(0)\}$, of the given functions. Since representing functions by their SRSFs loses
this initial value, this variable is treated separately. That is, a functional variable $f$ is analyzed
using the pair $(q, f(0))$ rather than just $q$. 
This way, the mapping from the function space $\fspace$ to $\ltwo \times \real$ is a bijection.
In practice, where $q$ is represented using a 
finite partition of $[0,1]$, say with cardinality $T$, the combined vector $h_i = [q_i~~f_i(0)]$ simply has dimension $(T+1)$ for fPCA. 
We can define a sample covariance operator for the aligned combined vector  $\tilde{h} = [\tilde{q}_1~~f_i(0)]$ as 
\begin{equation}
	K_h(s,t) = \frac{1}{n-1}\sum_{i=1}^n E[(\tilde{h}_i(s)-\mu_h(s))(\tilde{h}_i(t)-\mu_h(t))] \ \ \ \ \in \real^{(T+1) \times (T+1)}\ ,
\end{equation}
where $\mu_h = [\mu_q~~ \bar{f}(0)]$. 
Taking the SVD, $K_h=U_h\Sigma_h V_h^\T$ we can calculate the directions of principle variability in the given SRSFs using 
the first $p\leq n$ columns of $U_h$ and can be converted back to the function space $\mathcal{F}$, via integration, 
for finding the principal components of the original functional data. 
Moreover, we can calculate the observed principal coefficients as $\inner{\tilde{h}_i}{U_{h,j}}$.

\begin{figure}[t]
	\centering
    \begin{tabular}{rrr}
    \includegraphics{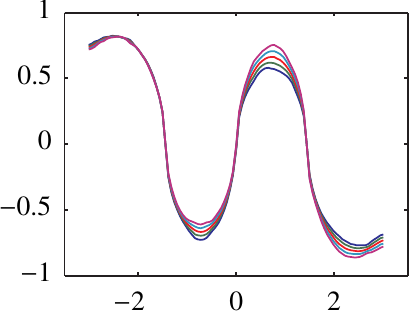}\hspace{.35em}
    \includegraphics{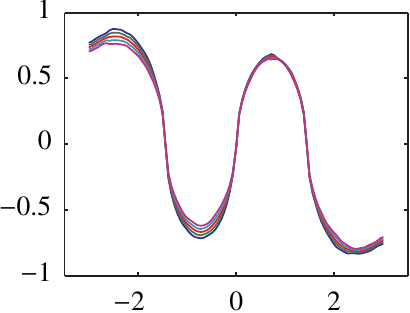}\hspace{.35em}
    \includegraphics{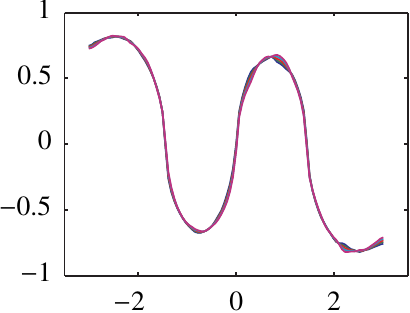}\\
		\includegraphics{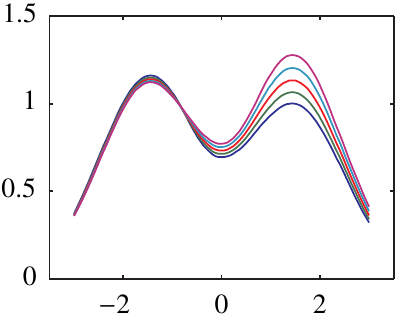}\hspace{.8em}
		\includegraphics{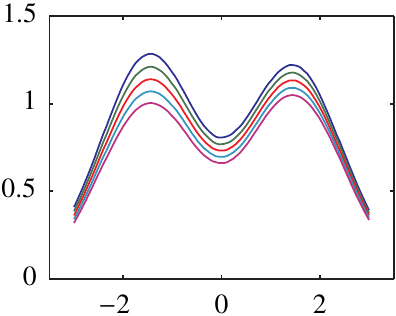}\hspace{.8em}
		\includegraphics{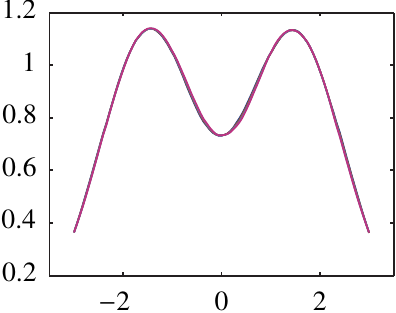}
    \end{tabular}
	\caption{Vertical fPCA of aligned functions in simulated data set of Fig.~\ref{fig:mean-result-sim}. The first row shows the main three principal directions in SRSF space and the second row shows the main three principal directions in function space.} 
	\label{fig:PCA-sim1} 
\end{figure}

One can then use this framework to visualize the vertical principal-geodesic paths.
The basic idea is to compute a few points along geodesic path $\tau \mapsto  \mu_h + \tau\sqrt{\Sigma_{h,jj}}U_{h,j}$ 
for $\tau \in \real$ in $\ltwo$, where $\Sigma_{h,jj}$ and $U_{h,j}$ are the $j^{th}$ singular value and column, respectively.  
Then, obtain  principle paths in the function space $\fspace$ by integration as described earlier. 
Figure \ref{fig:PCA-sim1} shows the results of vertical fPCA on the simulated data set from Fig.~\ref{fig:mean-result-sim}. 
It plots the vertical principal-geodesic paths of the SRSFs, $q_{\tau,j}$ for $\tau=-2,-1,0,1,2$ and $j=1,2,3$ and the vertical principal-geodesic paths in function space.
The first 3 singular values for the data are: $0.0481$, $0.0307$, and $0.0055$ with the rest being negligibly small.
The first principal direction corresponds to the height variation of the second peak while the second principal 
component captures the height variation of the first peak. 
The third principal direction has negligible variability.

\subsection{Modeling of Phase and Amplitude Components} \label{sec:models}
To develop statistical models for capturing the 
phase and amplitude variability, there are several possibilities. 
Once we have obtained the fPCA coefficients for these components we can impose probability on the 
coefficients and induce a distribution on the function space $\F$. Here we explore
two possibilities: a joint Gaussian model
and a non-parametric model.

Let $c = (c_1, \dots, c_{k_1})$ and $z = (z_{1},\dots, z_{k_2})$ be the dominant principal coefficients of the amplitude- 
and phase-components, respectively, 
as described in the previous two sections. Recall that $c_j = \inner{\tilde{h}}{U_{h,j}}$ and $z_j = \inner{v}{U_{\psi,j}}$.
We can reconstruct the amplitude component using $q = \mu_q + \sum_{j=1}^{k_1} c_j U_{h,j}$ and $f^s(t) = f^s(0) + \int_{0}^t q(s) |q(s)| ds$. 
Here, $f^s(0)$ is a random initial value. 
Similarly, we can reconstruct the phase component (a warping function) 
using $v = \sum_{j=1}^{k_2} z_j U_{\psi,j}$ and then using $\psi = \cos(\| {v}\|)\mu_{\psi} + \sin(\|{v}\|)\frac{ {v}}{\|{v}\|}$, 
and $\gamma^s(t) = \int_0^t \psi(s)^2 ds$. 
Combining the two random quantities, we obtain a random function $f^s\circ\gamma^s$.

\subsubsection{Gaussian Models on fPCA Coefficients}\label{sec:Gaussian}
In this setup the model specification reduces to the choice of models for $f^s(0)$, $c$, and $z$. 
We are going to model them as multivariate normal random variables. 
The mean of $f^s(0)$ is $\bar{f}(0)$ while the means of $c$ and $z$ are zero vectors. 
Their joint covariance matrix is of the type:
$
\left[ \begin{array}{ccc}
\sigma_0^2  & L_1 & L_2 \\
L_1^\T & \Sigma_h &  S \\
L_2^\T & S & \Sigma_{\psi} \end{array} \right] \in \real^{(k_1 + k_2 + 1) \times (k_1 + k_2 + 1)}
$.
Here, $L_1 \in \real^{1 \times k_1}$ captures the covariance between $f(0)$ and $c$, $L_2 \in \real^{1 \times k_2}$ between 
$f(0)$ and $z$, and $S \in \real^{k_1 \times k_2}$ between $c$ and $z$. 
As discussed in the previous sections $\Sigma_h \in \real^{k_1 \times k_1}$ and $\Sigma_{\psi} \in \real^{k_2 \times k_2}$ are diagonal matrices and are estimated directly from the data. We will call this resulting probability  model on the fPCA coefficients as $p_{Gauss}$. 

\subsubsection{Non-parametric Models on fPCA Coefficients}\label{sec:kde}
An alternative to the Gaussian assumption made above is the use of kernel density estimation \cite{book:silverman}, where the density of $f^s(0)$, each of the $k_1$ components of $c$, and the $k_2$ components of $z$ can be estimated using
\[p_{ker}(x) = \frac{1}{nb}\sum_{i=1}^n \mathcal{K}\left(\frac{x-x_i}{b}\right)\]
where $\mathcal{K}(\cdot)$ is the smoothing kernel, which is a symmetric function that integrates to 1, and $b>0$ is the smoothing parameter or bandwidth. 
A range of kernel functions can be used, but a common choice is the Gaussian kernel.

\section{Modeling Results} \label{sec:modelingresult}
\begin{figure}[htbp]
	\centering
  \begin{tabular}{cccc}
		\includegraphics{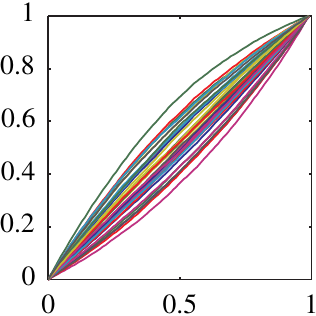}\hspace{.15em}
		\includegraphics{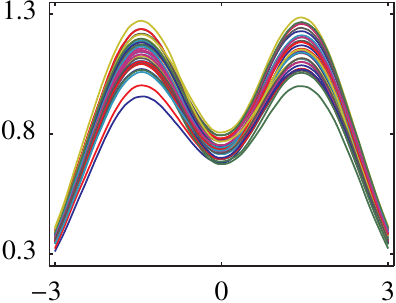}\hspace{.15em}
		\includegraphics{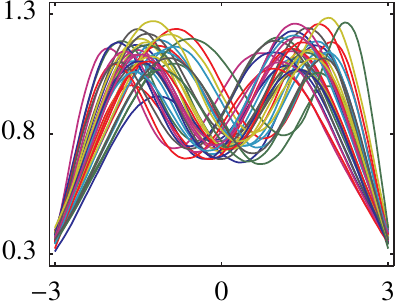}\hspace{.15em}
		\includegraphics{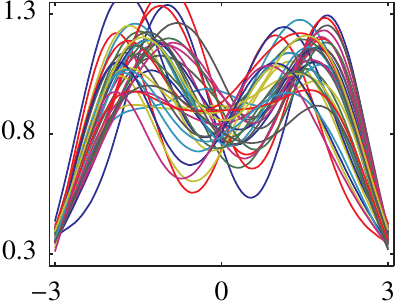}
    \end{tabular}
	\caption{Random samples from jointly Gaussian models on fPCA coefficients of $\gamma^s$ (left) and $f^s$ (middle), and their combinations $f^s \circ \gamma^s$ (right) for Simulated Data 1. The last plot are random samples if a Gaussian model is imposed on $f$ directly without any phase and 
	amplitude separation.}
	\label{fig:simu-sampling}
\end{figure}

We will now evaluate the models introduced in the previous section using random sampling. 
We will first estimate the means and the covariances from the given data, estimate the model parameters, and then generate random samples based on these estimated models. 
We demonstrate results on two simulated data sets used in Figs.~\ref{fig:mean-result-sim} and \ref{fig:alignresults} and one real data set being the Berkeley growth data\footnote{http://www.psych.mcgill.ca/faculty/ramsay/datasets.html}.
For the first simulated data set, 
shown in Fig.~\ref{fig:mean-result-sim}, 
we randomly generate 35 functions from the amplitude model and 35 domain-warping functions from the phase model and then combine them to generate random functions.
The corresponding results are shown in Fig.~\ref{fig:simu-sampling}, where the first panel is a set of random warping functions, the second panel is a set of corresponding amplitude functions, and the third panel shows their compositions. 
Comparing them with the original datasets (Fig.~\ref{fig:mean-result-sim}) we conclude that the random samples are very similar to the original data and, at least under a visual inspection, the proposed models are successful in capturing the variability in the given data.
Furthermore, if we compare these sampling results to the fPCA-based Gaussian 
model directly on $f$ (without separating the phase and amplitude components) in the last panel of Fig.~\ref{fig:simu-sampling}, 
we notice that our model is more consistent with the original data.  
A good portion of the samples from the non-separated model 
just contain three peaks or have a higher variation than the original data and some barely represent the original data. 

\begin{figure}[htbp]
	\centering
    \begin{tabular}{ccc}
		\includegraphics{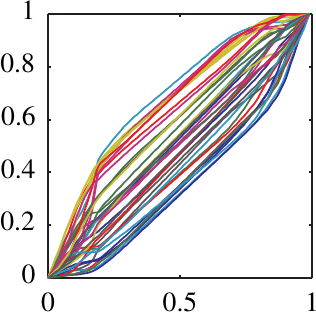}\hspace{.15em}
		\includegraphics{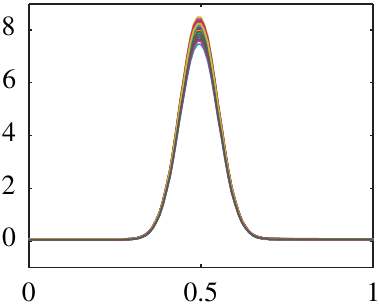}\hspace{.15em}
		\includegraphics{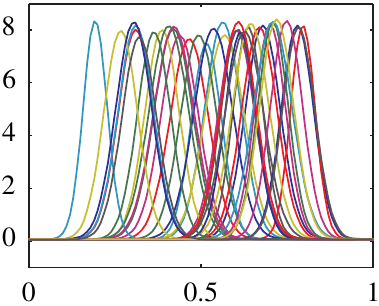}\hspace{.15em}
		\includegraphics{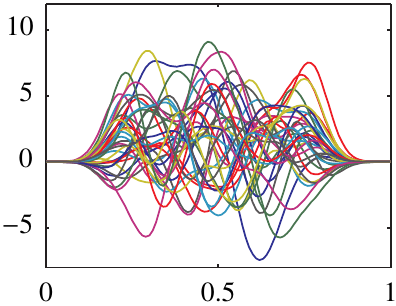}    
    \end{tabular}
	\caption{Random samples from jointly Gaussian models on fPCA coefficients of $\gamma^s$ (left) and $f^s$ (middle), and their combinations $f^s \circ \gamma^s$ (right) for Simulated Data 2. The last panel shows the random samples resulting from a Gaussian model imposed on $f$ directly.}
	\label{fig:toy-sampling}
\end{figure}

For the second simulated data set we use the data shown in Fig.~\ref{fig:alignresults} and perform vertical and horizontal fPCA.
As before, we randomly generate 35 functions from the amplitude model and 35 domain-warping functions from the phase model and then combine them to generate random functions. 
The corresponding results are shown row of Fig.~\ref{fig:toy-sampling}, where the first panel is a set of random warping functions, the second panel is a set of corresponding amplitude functions, and the last panel shows their compositions. 
Comparing them with the original data in Fig.~\ref{fig:alignresults} we conclude that the random samples are very similar to the original data and, under visual inspection, the proposed models are successful in capturing the variability in the given data. In this example performing fPCA directly on the function space does not correctly capture the data and fails to generate any single unimodal function shown in the last panel.

\begin{figure}[htbp]
	\centering
    \begin{tabular}{cccc}
		\includegraphics{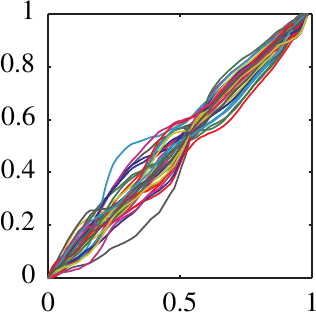}\hspace{.15em}
		\includegraphics{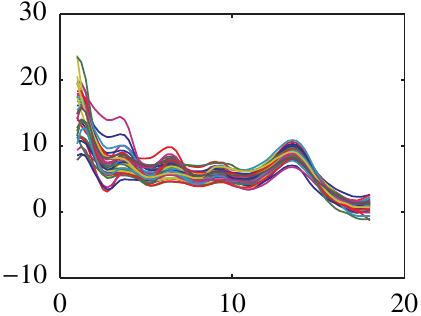}\hspace{.15em}
		\includegraphics{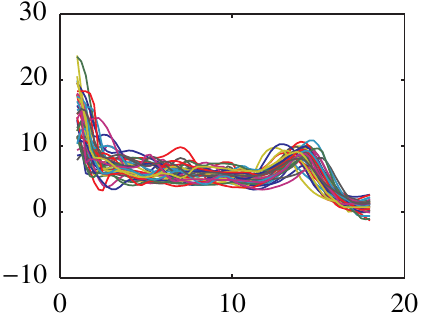}
		\includegraphics{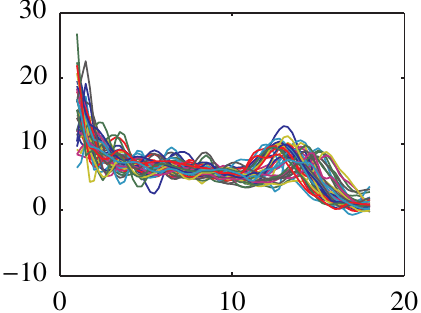}
    \end{tabular}
	\caption{From left to right: Random samples from jointly Gaussian models on fPCA coefficients of $\gamma^s$ and $f^s$, 
	respectively, and their combinations $f^s \circ \gamma^s$ for the Berkley Growth Data. The last panel shows the original data used in this experiment.}
	\label{fig:growth-sampling}
\end{figure}

For the Berkley growth data we again develop our phase and amplitude models then randomly generate 35 functions from the amplitude model and 35 domain-warping functions from the phase model. 
Then combine them to generate random functions. 
The corresponding results are shown row of Fig.~\ref{fig:growth-sampling}, where the first panel is a set of random warping functions, the second panel is a set of corresponding amplitude functions, and the last panel shows their compositions. 
Comparing them with the original data set in the last panel we conclude that the random samples are similar to the original data and, at least under a visual inspection, the proposed models are successful in capturing the variability in the given data. 

\section{Classification Using Phase and Amplitude Models}\label{sec:classification}
An important use of statistical models of functional data is in classification of future 
data into pre-determined categories. Since we have developed models for both 
amplitude and phase, one or both can be used for classification and analyzed for their 
performance.  Here we use a classical setup: 
a part of the data is used for training and estimation of model parameters while the remaining 
part is used for testing. This partition is often random and repeated many times to obtain 
an average classification performance. 
\\

\noindent {\bf Amplitude-Based Classification}:
As described earlier, we can impose a probability model for the amplitude data using the principal subspace associated
with the aligned SRSFs. The actual model is imposed on the principal coefficients $(c_1, c_2, \dots,c_{k_1})$, with 
respect to the basis $U_{h,1}, U_{h,2},\dots,U_{h,{k_1}}$. These basis elements, in turn, are determined using the 
training data. 
We can select a $k_1$ such that the cumulative energy $\sum_{j=1}^{k_1} \Sigma_{h,jj}/\sum_{j=1}^{T+1} \Sigma_{h,jj}$ is above a certain threshold, e.g., 90 percent. 
There are two choices of models: Gaussian and kernel-density estimator. 
Classification is performed by constructing the appropriate models for each class $C_1,\cdots,C_L$ of the data. 
Then, for a test sample $\tilde{h}_j \in \real^{T+1}$ project it to the principal 
subspace using an orthonormal basis $U_{hl} \in \real^{(T+1) \times k_1}$, one for each class, and calculate 
the likelihood under each class. The model with the largest likelihood represents the class assigned to $\tilde{h}_j$. 
Therefore, our classification rule is:
\begin{equation}
	\text{classify}(\tilde{h}_j) = \argmax_{C_l}~p_a(U^\T_{hl}\tilde{h}_j|K_{hl},\mu_{hl})\ \ ,\ \ \mbox{where}\ \ p_a = p_{Gauss}\ \ \mbox{or}\ \ p_{ker}\ .
	\label{eq:like_h} 
\end{equation}

\noindent {\bf Phase-Based Classification}:
Similarly, for the phase components, we can represent the shooting vectors, $\{v_i\}$, in a lower order dimensional space
using the first $k_2$ columns of $U_{\psi}$.
Where $k_2$ can be chosen similar to $k_1$ as described above.  
Once again, we can either define a Gaussian model or a kernel density 
estimator on these principal coefficients.
We can estimate the model parameters for each class $C_1,\cdots,C_L$ using the training data. 
Then, for a test sample's shooting vector $v_j$, we project it to each model's subspace and calculate the likelihood 
of $v_j$ under each pre-determined class. 
Therefore, our classification rule is:
\begin{equation}
	\text{classify}(v_j) = \argmax_{C_l}~p_{\psi}(U^\T_{\psi l}v_j|K_{\psi l})\  \ \mbox{where}\ \ p_{\psi} = p_{Gauss}\ \ \mbox{or}\ \ p_{ker}\ .
	\label{eq:like_v} 
\end{equation}

\noindent {\bf Joint Classification}:
Assuming independence we can combine the amplitude and phase classification rules as,
\begin{equation}
	\text{classify}(\tilde{h}_j,v_j) = \argmax_{C_l}~p_a(U^\T_{hl}\tilde{h}_j|K_{hl},\mu_{hl})p_{\psi}(U^\T_{\psi l}v_j|K_{\psi l}) 		
	\label{eq:comblike} 
\end{equation}
and classification is as described previously. \\

In this section, we present the classification results on a signature data \cite{Yeung-et-al:2004}, 
an iPhone-generated action data set from \cite{mccall-reddy-shah}, and a SONAR data set using models developed using vertical and horizontal fPCA.

\subsection{Signature Data}
\begin{figure}[htbp] 
	\centering 
	\begin{tabular}{cccc}
    \includegraphics{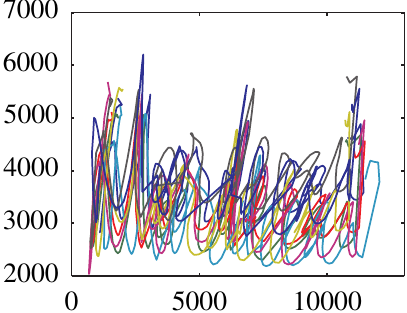}\hspace{.15em}
		\includegraphics{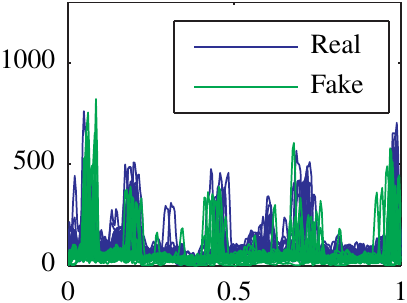}\hspace{.15em}
		\includegraphics{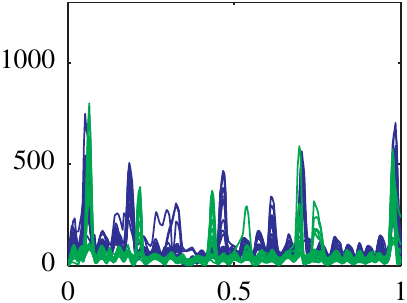}\hspace{.15em}
		\includegraphics{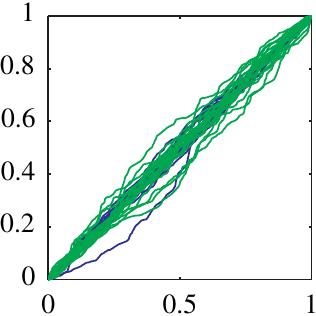}
	\end{tabular}
	\caption{From left to right: the original signature samples for one of the subjects, the corresponding tangential acceleration functions for both the real and fake signatures, the corresponding aligned functions, and warping functions.} 
	\label{fig:sigdata} 
\end{figure}

In this section, we test our classification method on a signature recognition data set from \cite{Yeung-et-al:2004}. 
The data was captured using a WACOM Intuos tablet. 
The data set consists of signature samples from 40 different subjects with 20 real signature samples of the subject and another 20 samples which are forgeries of the subject's signature.
In our analysis we are going to distinguish between the real and forgery signature for two of the subjects using the tangential acceleration.
The tangential acceleration is computed as $A(t) = \sqrt{[X''(t)]^2+[Y''(t)]^2}$.
To have a robust estimate of the SRSF $\{q_i\}$, we first smooth the original functions 100 times $\{f_i\}$ using a standard box filter $[1/4, 1/2, 1/4]$. 
That is, numerically we update the signals at each discrete point by $f_i(x_k) \rightarrow \left( \frac{1}{4} f_i(x_{k-1}) 
+\frac{1}{2} f_i(x_k) + \frac{1}{4} f_i(x_{k+1}) \right)$.
The smoothed acceleration functions are aligned in each class (real vs. fake) using our alignment algorithm from Section~\ref{sec:theory}.
An example signature with 10 realizations is shown in Fig.~\ref{fig:sigdata} along with the corresponding acceleration functions for both the real and fake signatures, the corresponding aligned functions, and warping functions.

Models were generated for the three classes as was outlined in Section~\ref{sec:models} by performing vertical and horizontal fPCA on the aligned data and the warping functions, respectively. We then impose a multivariate Gaussian model, $p_{Gauss}$, on the reduced data for each class, it is assumed here that 
the cross-covariances $L_1$ and $L_2$ are zero.
The threshold to select the number of dimensions, $k_1$ and $k_2$, was set at 95\%. 
Classification for the amplitude component only was performed as described in Section~\ref{sec:classification} using the classification rule in (Eqn. \ref{eq:like_h}) and was evaluated using 5-fold cross-validation. 
Similarly, the classification rule in (Eqn. \ref{eq:like_v}) were used for the phase component. 
Moreover, the joint classification was performed using (Eqn. \ref{eq:comblike}).
Table~\ref{tab:sigkfoldratesU1a} presents the the mean and standard deviation (shown in parentheses) of the classification rates from the cross-validation for the three rules.
As well as comparing to the standard $\ltwo$ where models were generated directly on the original data, dimension reduction with fPCA, and imposing a multivariate normal distribution.
Corresponding results for another subject, U13 is presented in Table~\ref{tab:sigkfoldratesU1b}.

\begin{table}[htbp]
	\begin{center}
  \subfloat[Subject U1]{
  \begin{tabular}{ c c c }
        & Gaussian & Kernel Density \\ \hline\hline
    amplitude only & 0.93 (0.07) & 0.78 (0.19) \\ \hline
    phase only & 0.65 (0.16) & 0.75 (0.09) \\ \hline
    phase and amplitude & 0.90 (0.05) & 0.80 (0.07) \\ \hline
    standard $\ltwo$ & 0.60 (0.14) & 0.55 (0.11) \\ \hline
  \end{tabular}
  \label{tab:sigkfoldratesU1a}}\hspace{1em}
  \subfloat[Subject U13]{
  \begin{tabular}{ c c c }
        & Gaussian & Kernel Density \\ \hline\hline
    amplitude only & 0.75 (0.14) & 0.78 (0.21) \\ \hline
    phase only & 0.50 (0.01) & 0.50 (0.01) \\ \hline
    phase and amplitude & 0.58 (0.11) & 0.60 (0.10) \\ \hline
    standard $\ltwo$ & 0.50 (0.01) & 0.53 (0.06) \\ \hline
  \end{tabular}
  \label{tab:sigkfoldratesU1b}}
	\end{center}
	\caption{Mean classification rate and standard deviation (in parentheses) for 5-fold cross-validation on the signature data.}
	\label{tab:sigkfoldratesU1}
\end{table}

The classification rates have a low standard deviation indicating good generalization, though we do have a little variation for the phase only model.
For both subjects the amplitude only rule greatly outperforms both the phase only rule and the standard $\ltwo$ with the best performance of 93\% and 75\% for subjects U1 and U13, respectively.
Since the phase only rule performs poorly combining it with the amplitude only rule brings down the overall performance.  
The alignment and modeling using a proper distance improves the overall classification performance of the data.
To compare the results between $p_{Gauss}$ and $p_{kern}$, we classified the data again forming models using $p_{kern}$ which was discussed in Section~\ref{sec:classification}, 
where each of the $k_1$ and $k_2$ components has an estimated density using a kernel density estimator and independence is assumed. 
We used the Gaussian kernel function and the bandwidth was selected automatically based upon the data using the method presented by \cite{botev-grotwoski-kroese}.
Classification using the three classification rules was performed using 5-fold cross-validation.
Table~\ref{tab:sigkfoldratesU1a} and Table~\ref{tab:sigkfoldratesU1b} present the the mean and standard deviation of the classification rates from the cross-validation for the three rules as well as comparing to the standard $\ltwo$. 
Models were generated directly on the original data using fPCA and the kernel density estimator for subjects U1 and U13, respectively.
We see an improvement in the phase only method for subject U1 and reduction in performance for the other methods, this suggest the warping functions have some non-Gaussian behavior.
However, for subject U13 there is a minimal change between the Gaussian and kernel estimator.

\subsection{iPhone Action Data}
\begin{figure}[t] 
  \centering 
  \begin{tabular}{rrr}
    \includegraphics{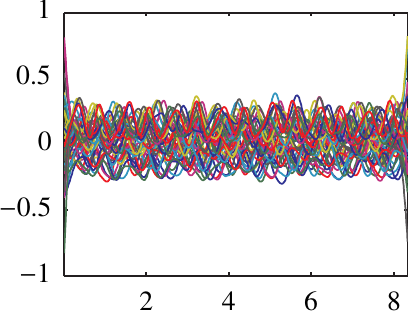}&
    \includegraphics{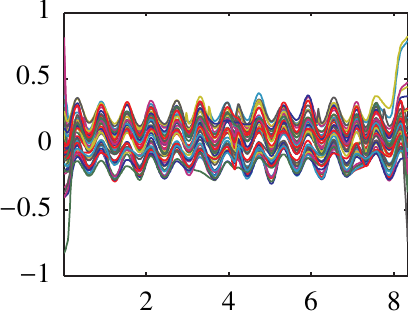}&
    \includegraphics{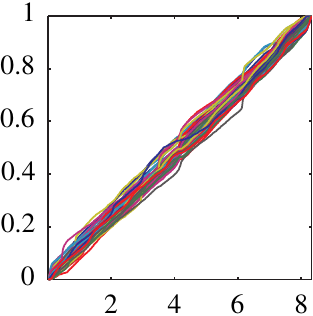}\\
    \includegraphics{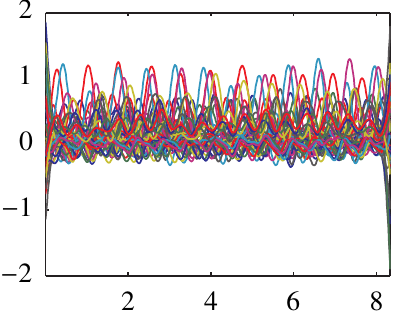}&
    \includegraphics{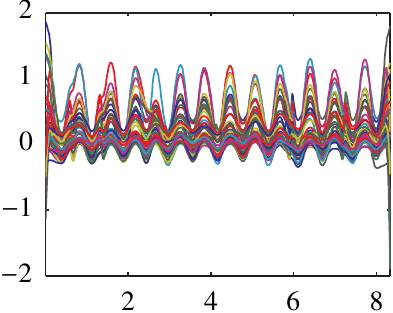}&
    \includegraphics{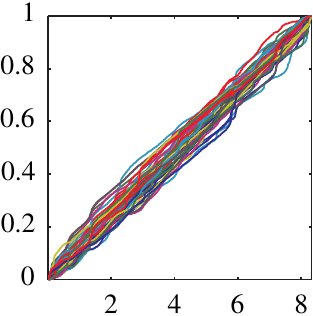}\\
    \includegraphics{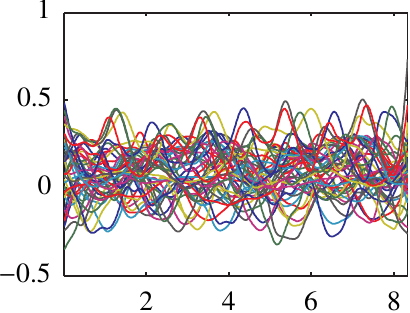}&
    \includegraphics{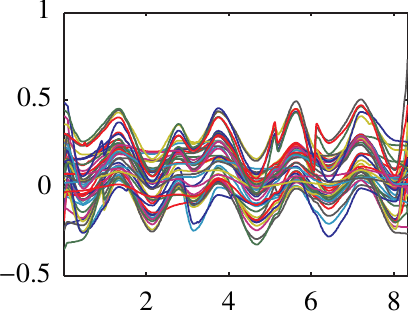}&
    \includegraphics{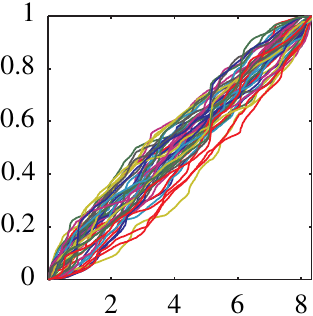}
  \end{tabular}
  \caption{Original iPhone functions for the walking, jumping, and climbing activities in the first column (in corresponding descending order) with the corresponding aligned functions and warping functions in the second and third columns, respectively.} 
  \label{fig:iphonedata} 
\end{figure}

This data set consists of aerobic actions recorded from subjects using the Inertial Measurement Unit (IMU) on an Apple iPhone 4 smartphone. 
The IMU includes a 3D accelerometer, gyroscope, and magnetometer. 
Each sample was taken at 60Hz, and manually trimmed to 500 samples (8.33s) to eliminate starting and stopping movements and the iPhone is always clipped to the belt on the right hand side of the subject. 
There is a total of 338 functions for each measurement on the IMU and the actions recorded consisted of biking, climbing, gym bike, jumping, running, standing, treadmill, and walking.
With the number of samples being 30, 45, 39, 45, 45, 45, 44, and 45, respectively for each action.
For more information on the data set the reader is referred to~\cite{mccall-reddy-shah}. 
For our experiments we used the accelerometer data in the $x$-direction.
Again, to have a robust estimate of the SRSF $\{q_i\}$, we first smooth the original signals 100 times $\{f_i\}$ using the standard box filter described in the previous section.
As with the previous data set, the smoothed iPhone data are aligned in each class (activity) using our method.
A selected subset of functions from three activities is shown in Fig.~\ref{fig:iphonedata} along with corresponding aligned functions and warping functions.

To perform the classification, models were generated for the 8 classes by performing vertical and horizontal fPCA on the aligned data and the warping functions then imposing a multivariate Gaussian on the reduced data for each class.
The threshold to select the number of dimensions, $k_1$ and $k_2$, was set at 95\%.
Classification was performed as in the previous section.
Table~\ref{tab:iphonekfoldrates} presents the mean and standard deviation of the classification rates for the cross-validation for all three rules as well as comparing to the standard $\ltwo$.
\begin{table}[htbp]
	\begin{center}
  \begin{tabular}{c c c}
        & Gaussian & Kernel Density \\ \hline\hline
    amplitude only & 0.60 (0.04) & 0.62 (0.05) \\ \hline
    phase only & 0.34 (0.06) & 0.35 (0.06) \\ \hline
    phase and amplitude & 0.62 (0.08) & 0.62 (0.07) \\ \hline
    standard $\ltwo$ & 0.12 (0.02) & 0.12 (0.02) \\ \hline
  \end{tabular}
	\end{center}
	\caption{Mean classification rate and standard deviation (in parentheses) for 5-fold cross-validation on the iPhone data.}
	\label{tab:iphonekfoldrates}
\end{table}
The classification rates have a low standard deviation indicating good generalization. 
The phase only rule and the amplitude only rule, drastically out perform the standard $\ltwo$ with the combination providing the best performance at 62\%. 
The alignment and modeling using a proper distance improves the overall classification performance of the data.
We again used the kernel density estimator to compare the results with the Gaussian assumption and the results are presented in
Table~\ref{tab:iphonekfoldrates}. 
Using the kernel density estimator we see only minor improvements in the phase only rule, suggesting the Gaussian assumption is sufficient for this data.

\subsection{SONAR Data}
The data set used in these experiments was collected at the Naval Surface Warfare Center Panama City Division (NSWC PCD) test pond. 
For a description of the pond and measurement setup the reader is referred to~\cite{art:kargl}. 
The raw SONAR data was collected using a 1 - 30$kHz$ LFM chirp and data was collected for nine proud targets that included a solid aluminum cylinder, an aluminum pipe, an inert 81$mm$ mortar (filled with cement), a solid steel artillery shell, two machined aluminum UXOs, a machined steel UXO, a de-militarized 152$mm$ TP-T round, a de-militarized 155$mm$ empty projectile (without fuse or lifting eye), and a small aluminum cylinder with a notch. 
The aluminum cylinder is 2$ft$ long with a 1$ft$ diameter; while the pipe is 2$ft$ long with an inner diameter of 1$ft$ and 3/8 inch wall thickness.

\begin{figure}[t] 
	\centering 
		\includegraphics{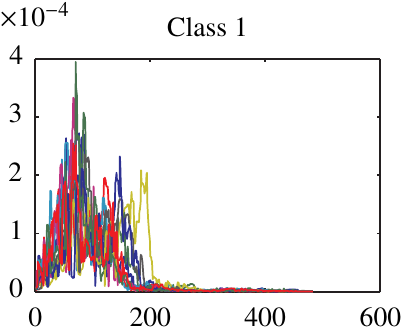}
		\includegraphics{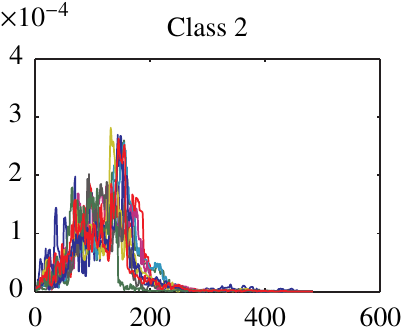}
		\includegraphics{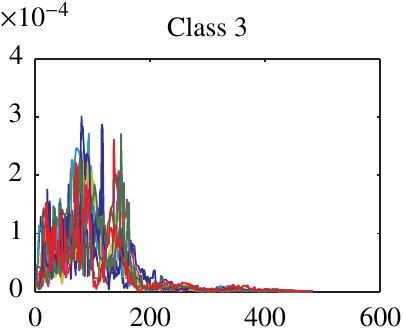}\\
		\includegraphics{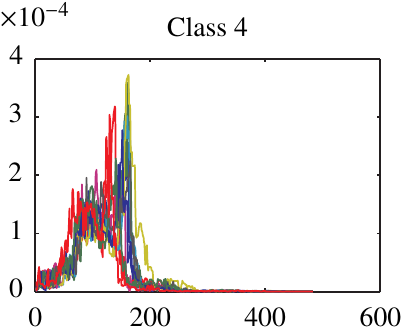}
		\includegraphics{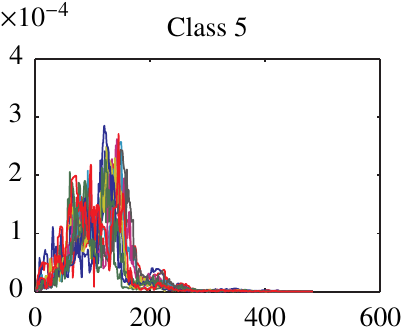}
		\includegraphics{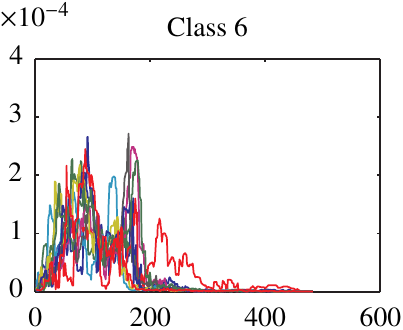}\\
		\includegraphics{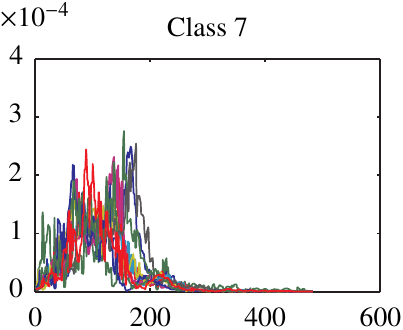}
		\includegraphics{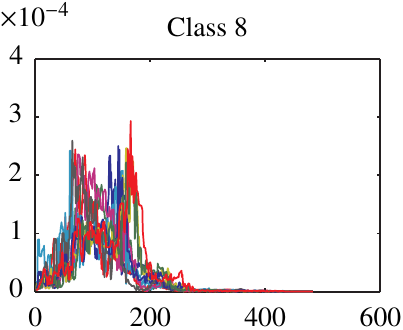}
		\includegraphics{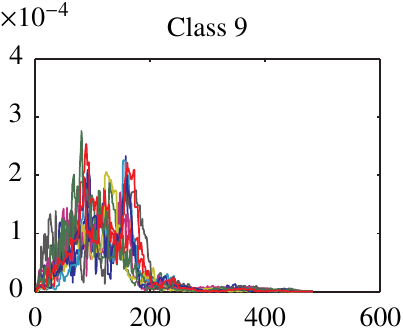}
	\caption{Original SONAR functions in each of the 9 classes.} 
	\label{fig:data} 
\end{figure}

The acoustic signals were generated from the raw SONAR data to construct target strength as a function of frequency and aspect angle. 
Due to the relatively small separation distances between the targets in the measurement setup, the scattered fields from the targets overlap. 
To generate the acoustic templates (i.e., target strength plot of frequency versus aspect), synthetic aperture sonar (SAS) images were formed and then an inverse imaging technique was used to isolate the response of an individual target and to suppress reverberation noise. 
A brief summary of this process is as follows: The raw SONAR data are matched filtered and the SAS image is formed using the $\omega-k$ beamformer \cite{book:soumekh2}. 
The target is then located in the SAS image and is windowed around selected location. 
This windowed image contains the information to reconstruct the frequency signals associated with a given target via inverting the $\omega-k$ beamformer \cite{art:khwaja} and the responses were aligned in rage using the known acquisition geometry. 
For the nine targets, 2000 different data collections runs were done, and 1102 acoustic color templates were generated using the method described above from the data set. 
From the acoustic color maps, one-dimensional functional data was generated by taking slices at aspect value of $0^\circ$ and therefore generating 1102 data samples.
We will apply our method to this SONAR data, where there are $n = 1102$ SONAR signals with nine target classes and the numbers of functions in the nine classes are $\{n_i\}_{i=1}^9 = \{131, 144, 118, 118, 121, 119, 120, 114, 117\}$ and are sampled using 483 points.
A selected subset of functions in each class from the original data is shown in Fig.~\ref{fig:data}. 
We observe that the original data are quite noisy, due to both the compositional and the additive noise, increasing variability within class and reducing separation across classes. 
This naturally complicates the task of target classification using SONAR signals. 

\begin{figure}[htb] 
  \centering 
    \includegraphics{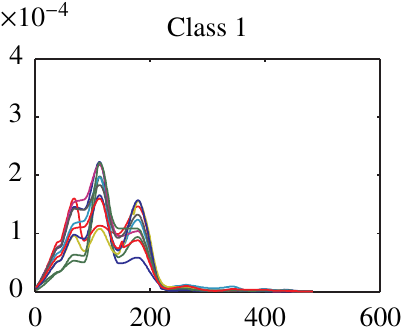}
    \includegraphics{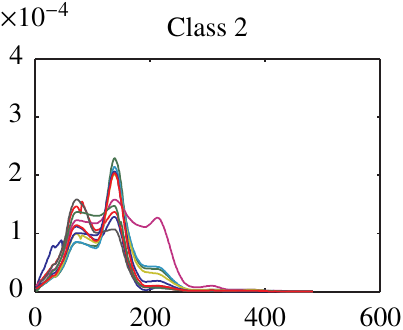}
    \includegraphics{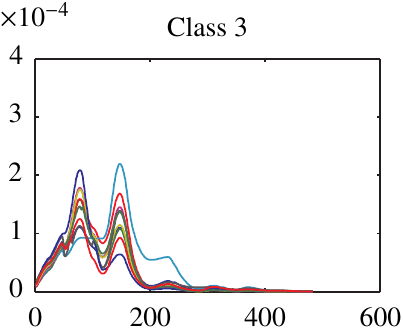}\\
    \includegraphics{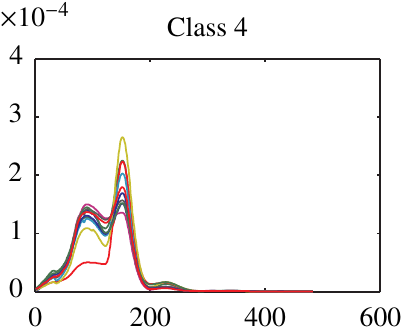}
    \includegraphics{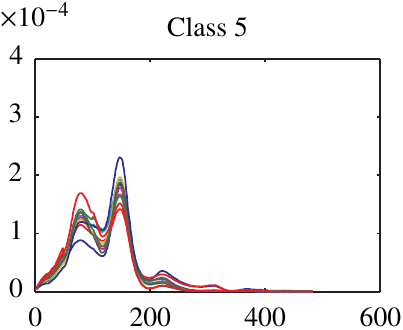}
    \includegraphics{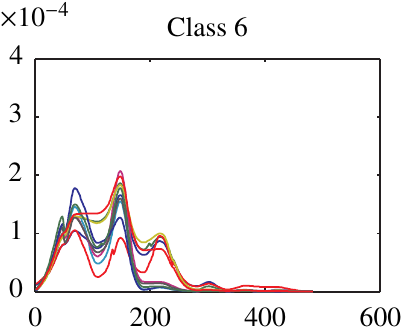}\\
    \includegraphics{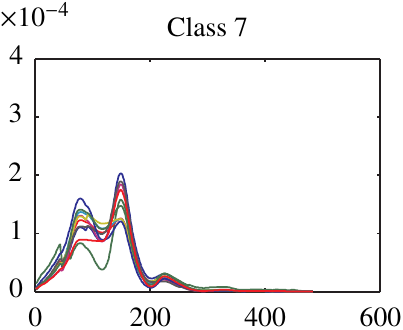}
    \includegraphics{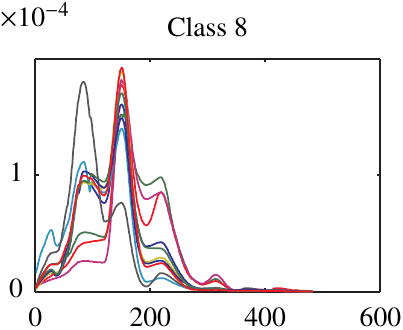}
    \includegraphics{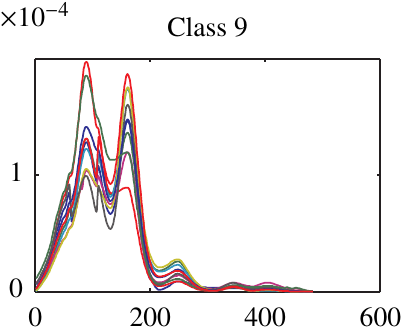}
  \caption{Aligned and Smoothed SONAR functions in each of the 9 classes.} 
  \label{fig:datafn} 
\end{figure}

To again have a robust estimate of the SRSF $\{q_i\}$, we first smooth the original signals 25 times $\{f_i\}$ using the standard box filter described previously.
As with the previous data sets, the smoothed SONAR data are aligned in each class using our method.
Models were generated for the three classes by performing vertical and horizontal fPCA on the aligned data and the warping functions then, imposing a multivariate Gaussian on the reduced data for each class, with the aligned data shown in Fig.~\ref{fig:datafn}.
The threshold to select the number of dimensions, $k_1$ and $k_2$, was set at 90\%. 
Table~\ref{tab:sonarkfoldrates} presents the classification rates for the cross-validation for all three rules as well as comparing to the standard $\ltwo$.

\begin{table}[htbp]
  \begin{center}
  \begin{tabular}{ c c c }
        & Gaussian & Kernel Density \\ \hline\hline
    amplitude only & 0.44 (0.03) & 0.47 (0.02) \\ \hline
    phase only & 0.42 (0.02) & 0.43 (0.02) \\ \hline
    phase and amplitude & 0.54 (0.03) & 0.53 (0.03) \\ \hline
    standard $\ltwo$ & 0.33 (0.01) & 0.34 (0.02) \\ \hline
  \end{tabular}
  \end{center}
  \caption{Mean classification rate and standard deviation (in parentheses) for 5-fold cross-validation on SONAR data.}
  \label{tab:sonarkfoldrates}
\end{table}

The classification rates have low standard deviation indicating good generalization for the SONAR data. 
The phase only rule and the amplitude only rule out perform the standard $\ltwo$ with the combination providing the best performance at 54\%. 
The alignment and modeling using a proper distance improves the overall classification performance of the data.
We again used the kernel density estimator to compare the results with the Gaussian assumption and the results are presented in
Table~\ref{tab:sonarkfoldrates}.
Using the kernel density estimator we see improvements in the classification results. 
However, nothing is a dramatic improvement suggesting the Gaussian assumption is sufficient for this data.

\section{Conclusions}\label{sec:conclusion}
The statistical modeling and classification of functional data with phase variability is a challenging and complicated task. 
We have proposed a comprehensive approach that solves the problem of registering and modeling functions in a 
joint, metric-based framework. 
The main idea is to use an elastic distance to separate the given functional data into phase and amplitude components, and to develop individual models for these components. 
The specific models suggested in this paper use fPCA and imposition of either multivariate Gaussian or nonparametric models on the coefficients. 
The strengths of these models are illustrated in two ways: random sampling and model-based classification of functional data. 
In the case of classification, we consider applications involving handwritten signatures, motion data collected using iPhones, and SONAR signals.
We illustrate the improvements in classification performance when the proposed models involving separate phase and amplitude components are used.

\section*{Acknowledgment}
This research was supported by the Naval Surface Warfare Center Panama City Division In-house Laboratory Independent Research program funded by the Office of Naval Research (ONR) and was also supported in part by the grants NSF DMS 0915003 and NSF DMS 1208959.
The authors would like to thank Dr. Frank Crosby and Dr. Quyen Huynh at NSWC PCD for their technical support during this work and the two anonymous referees and Associate Editor for their constructive comments and suggestions, which led to a significant improvement of the paper.

\section*{References}
\bibliography{JDTBib} 
\bibliographystyle{elsarticle-harv}

\end{document}